\documentclass[preprint,aps]{revtex4}

\usepackage{amssymb}
\usepackage{graphicx}
\usepackage{dcolumn}
\usepackage{bm}
\usepackage{amsmath}
\usepackage{bibentry}
\usepackage{natbib}

\begin{document}

\title{Nonlinear modal coupling in a high-stress doubly-clamped
nanomechanical resonator}
\author{K. J. Lulla}
\altaffiliation[Current address: ]{Institut N\'eel, CNRS et Universit\'e Joseph Fourier, BP 166, 38042 Grenoble
Cedex 9, France}
\author{R. B. Cousins}
\author{A. Venkatesan}
\altaffiliation[Current address: ]{IISER, Mohali, Knowledge city, Sector 81, SAS Nagar, Manauli PO 140306, India}
\author{M. J. Patton}
\author{A. D. Armour}
\author{C. J. Mellor}
\author{J. R. Owers-Bradley}
\affiliation{School of Physics and Astronomy, University of Nottingham, Nottingham NG7
2RD, United Kingdom}
\date{\today }

\begin{abstract}
We present results from a study of the nonlinear intermodal coupling between
different flexural vibrational modes of a single high-stress, doubly-clamped
silicon nitride nanomechanical beam. The measurements were carried out at
100 mK and the beam was actuated using the magnetomotive technique. We
observed the nonlinear behavior of the modes individually and also measured
the coupling between them by driving the beam at multiple frequencies. We
demonstrate that the different modes of the resonator are coupled to each
other by the displacement induced tension in the beam, which also leads to
the well known Duffing nonlinearity in doubly-clamped beams.
\end{abstract}

\maketitle



\section{Introduction}

The nonlinear dynamics of nanoelectromechanical systems (NEMS) has attracted
considerable interest over recent years.\cite%
{Blick2002,Lifshitz,Rhoads,Postma2005,aldridge,kozinsky2,buks2,kozinsky,Karabalin2009,Westra2010,Dunn2010,collin}
Nonlinear behavior is of practical importance in NEMS as it can limit their
use as sensors operating in the linear regime,\cite{ekinci,Postma2005}
though it also opens up the possibility of devices that are designed to
exploit it.\cite{aldridge,guerra} However, the very high operating
frequencies of NEMS devices (which typically have resonant frequencies in
the MHz range) make them ideal for investigating fundamental aspects of
nonlinear dynamics.\cite%
{aldridge,kozinsky,buks2,Karabalin2009,Westra2010,collin} Because NEMS
devices are engineered systems and their nonlinearities can be extensively
tuned,\cite{kozinsky2} it should be possible to use them to study complex
nonlinear phenomena\cite{cross} in which a large number of different
mechanical modes interact.\cite{buksmems} Furthermore, when cooled to very
low temperatures, the nonlinear behavior in nanomechanical devices would
provide important signatures of the transition from classical to quantum
regimes.\cite{quantum}

Nonlinearity can arise in a wide variety of different ways in NEMS devices.%
\cite{Lifshitz} External potentials generated electrostatically by fixed
electrodes can be used to induce (or suppress) nonlinear behavior in NEMS.%
\cite{kozinsky2} However, nonlinearities can be intrinsic to a given device
or material; for example when a doubly clamped beam is excited its
displacement leads to a stretching of the beam which generates a Duffing
type nonlinearity in a given mode as well as generating inter-mode couplings.%
\cite{Westra2010,Dunn2010} The damping of nanomechanical devices can also be
quite nonlinear, something which appears to be particularly important for
carbon based devices.\cite{eichler}

Experiments have begun to explore the coupled nonlinear dynamics of
mechanical modes either in two adjacent resonators\cite{Karabalin2009} or
occuring as different harmonics within a single beam.\cite%
{Westra2010,Dunn2010} In a recent work Westra et al.\cite{Westra2010}
explored the nonlinear coupling between the first and third harmonics of a
micromechanical doubly clamped beam with mode frequencies in the kHz range.
Although the Q-factor of the modes was rather low ($\simeq 10^2$) they were
able to see clear evidence of nonlinear coupling between the modes induced
by the stretching nonlinearity in their beam.

In this paper, we present a systematic study of the nonlinear response of
three of the flexural modes of a nanomechanical beam resonator, together
with the associated inter-mode couplings. We report the results of a series
of measurements taken at 100\,mK on the fundamental, third and fifth
harmonics (all of which are in the MHz range) of a doubly-clamped beam
fabricated from high-stress ($\sim$ 1 GPa) silicon nitride. In a first set
of measurements, we excite one mode at a time and find that they display a
Duffing like behavior which can be understood quite naturally in terms of
the nonlinearity arising from the stretching of the beam. We then measured
the response when two modes are driven at the same time, with one mode
acting as a probe of the displacement of the other. The high tensile stress
and the low temperature mean that the Q-factor of the modes is high enough%
\cite{sini} to allow us to observe the higher mode resonance signals as well
as the frequency shifts induced by nonlinear mode-mode couplings which are
much larger than the linewidth. Using the single-mode results to calibrate
our measuring scheme we are able to make a detailed quantitative comparison
(without the need for any fitting parameters) between our results and the
predictions of a simple model of the stretching nonlinearity.

This paper is organized as follows. Section II introduces a simple
theoretical model of the stretching nonlinearity in beam resonators. The
experimental set-up is outlined in Sec.\ III and in Sec.\ IV we describe our
results. We give our conclusions in Sec.\ V.

\section{Theoretical background}

Here we review the theoretical description  of a beam under tension,\cite%
{Bokaian} starting from Euler-Bernoulli equation extended to include the
geometric nonlinearity arising from the stretching of the beam that
accompanies its flexure.\cite{Nayfeh,Lifshitz} For a single mode, this
nonlinearity leads naturally to the Duffing equation where a term
proportional to the cube of the displacement is added to the usual harmonic
equation of motion.\cite{Lifshitz} However, the non-linearity also generates
important couplings between different modes.\cite{Westra2010}

We start by considering a doubly-clamped beam of length $L$, lying along the
$x$-axis, with cross-sectional area $A=wh$, where $w$ is the width and $h$
the thickness. The beam is assumed to be under intrinsic tension, $T_{0}$,
and we include, to lowest order, the nonlinear tension arising from the
stretching of the beam.\cite{Lifshitz} The equation of motion for the
displacement, $y$, is given by
\begin{equation}
\rho A\ddot{y}+\eta \dot{y}+EI_{y}y^{\prime \prime \prime \prime }-\left[
T_{0}+\frac{EA}{2L}\int_{0}^{L}(y^{\prime })^2dx\right] y^{\prime \prime
}=F_{L},  \label{eq:waveqtension}
\end{equation}%
where $\rho$ is the density, $\eta$ characterizes the damping, $I_{y}=hw^3/12
$ is the moment of inertia, $F_{L}$ is the force per unit length exerted on
the beam and $E$ the Young's modulus. The non-linearity arises from the
second term within the square brackets.

The mode frequencies and mode functions for the corresponding linear problem
are known\cite{Bokaian} and provide a convenient starting point for
describing the non-linear regime. If the system is driven harmonically, $%
F_L=F_n\cos(\omega_n t)$, at a frequency close to one of the linear mode
frequencies $\omega_{n,0}$, we can then approximate the displacement of the
beam as $y(x,t)=wg_n(x)u_n(t)$ where $g_n(x)$ is the $n$-th mode function,
normalized so that $\int_0^1 g_n^2(\tilde{x})d\tilde{x}=1$ where $\tilde{x}%
=x/L$.

The dimensionless function $u_n(t)$ obeys the equation of motion,\cite%
{Lifshitz}
\begin{equation}
\frac{d^2u_{n}}{dt^2}+\frac{\omega_{n,0}}{Q_{n}}\frac{d u_{n}}{dt}%
+\omega_{n,0}^2{u}_{n}+\lambda I_{nn}^{2}{u}_{n}^{3}=f_{n}\cos ({\omega}%
_{n}t),  \label{eq:drivensingle}
\end{equation}
where $\lambda =(Ew^2)/(2\rho L^{4})$, $f_n=F_n\xi_n/(hw^2\rho)$, with $%
\xi_n=\int_0^1d\tilde{x}g_n(\tilde{x})$ (the mode parameter), and the
Q-factor of the $n$-th mode is given by $Q_n=\omega_{n,0}(A\rho/\eta)$. The
strength of the nonlinearity is controlled by the parameter $I_{nn}$ which
is defined by an integral of the form, $I_{ij}=\int_0^1d\tilde{x}g_i^{\prime
}(\tilde{x})g_j^{\prime }(\tilde{x})$ (with $i$ and $j$ labeling any two
modes of the beam).

Since the equation of motion [Eq.\ \ref{eq:drivensingle}] is that of the
familiar Duffing oscillator, we proceed to analyse it using standard methods.%
\cite{Hand} Substituting a solution of the form,
\begin{equation}
{u}_{n}=a_{n}\cos ({\omega}_{n}t)+b_{n}\sin ({\omega}_{n}t),
\label{eq:drivensol}
\end{equation}%
into Eq.\ \ref{eq:drivensingle}, we obtain an equation for the amplitude of
the motion, $r_n=(a_n^2+b_n^2)^{1/2}$,
\begin{equation}
r_{n}=f_{n}\left[ \left(\omega_{n,0}^2+\frac{3}{4} \lambda I_{nn}^{2}
r_{n}^{2}-{\omega}_{n}^{2}\right)^{2}+\left( \frac{{\omega}_{n}{\omega}_{n,0}%
}{Q_{n}}\right) ^{2}\right]^{-1/2}.  \label{eq:rn}
\end{equation}
For a large enough quality factor, $Q_n$, and for relatively small
amplitudes, $r_{n}$, the main effect of the nonlinearity is to shift the
frequency of peak response to
\begin{equation}
\omega_{n,1}=\omega_{n,0}\sqrt{1 + \frac{3}{4}\lambda I_{nn}^{2} r_{n}^{2}}.
\label{eq:singlefreq}
\end{equation}
In the weakly nonlinear regime, the frequency of the resonator shifts from
its natural (undriven) value quadratically with increasing amplitude
(frequency pulling).\cite{Hand,Nayfeh}

We now consider a situation where the drive contains two harmonic
components,
\begin{equation}
F_L=F_n\cos(\omega_n t)+F_m\cos(\omega_m t),
\end{equation}
with the two drives frequencies $\omega_n$ and $\omega_m$ chosen to be close
to the resonant frequencies of two different modes, $\omega_{0n}$ and $%
\omega_{0m}$, respectively. In this case we assume the beam displacement can
be approximated as, $y(x,t)=w[u_n(t)g_n(x)+u_m(t)g_m(x)]$.

Following the same approach as for the one mode case, assuming $u_n$ and $u_m
$ each oscillate at a single frequency (i.e.\ they take the form of Eq.\ \ref%
{eq:drivensol}), we obtain a modified expression for the amplitude of mode $n
$,
\begin{equation}
r_n=f_n\left[ \left(\omega_{n,2}^2-{\omega}_{n}^{2}\right)^{2}+\left( \frac{{%
\omega}_{n}{\omega}_{n,0}}{Q_{n}}\right) ^{2}\right]^{-1/2}.  \label{eq:rn2}
\end{equation}
where
\begin{equation}
\omega_{n,2}=\omega_{n,0}\sqrt{ 1+ \frac{3}{4}\lambda I_{nn}^{2} r_{n}^{2}+{\lambda}\left(\frac{1}{2}%
I_{nn}I_{mm}+I_{nm}^2\right)r_m^2.}  \label{eq:twofreq}
\end{equation}
Thus we see that the frequency of a given mode depends not just on the
amplitude of its own motion, but the amplitude of the other mode that is
excited. In particular, the frequency shift of one mode initially grows
quadratically with the amplitude of the other one.

\section{Experimental Set-up}

\begin{figure}[htbp]
\centering
\includegraphics{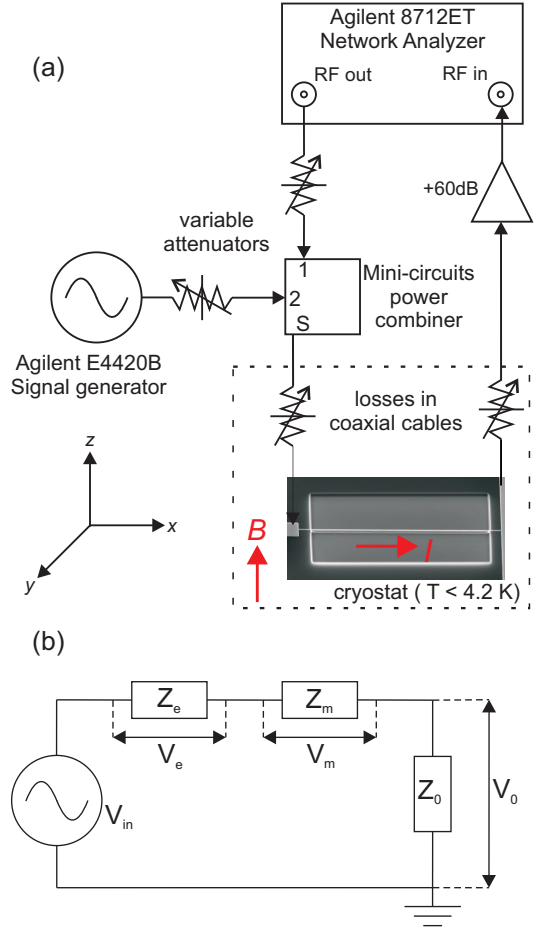}
\caption{(a) Schematic diagram of the measurement set-up. The different
modes of the beam were driven with two separate signal generators. A voltage
signal proportional to the emf generated by the beam was detected at the
input of the network analyzer after passing through a pre-amplifier. (b)
Circuit model for transmission measurements, with external circuit
impedances $Z_e$ and $Z_0$, sample impedance $Z_m$, the drive voltage $V_{%
\mathrm{in}}$ and the measured voltage $V_0$.}
\label{fig:fig1}
\end{figure}

\begin{figure}[htbp]
\centering
\includegraphics[width=8cm]{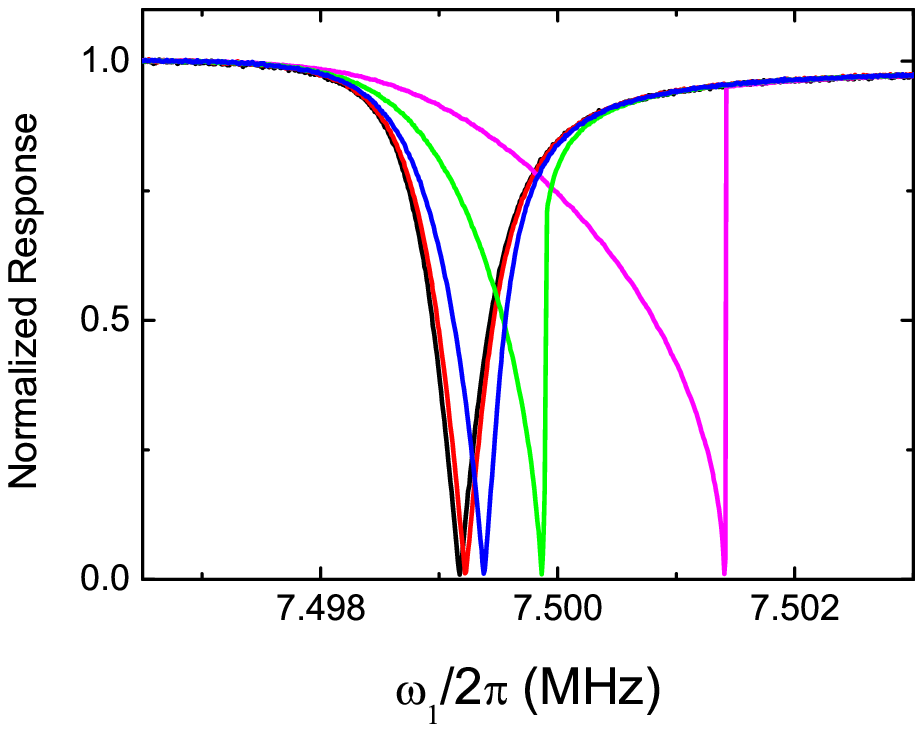}
\caption{Resonant response of mode 1 (7.5\,MHz). The curves from left to
right correspond to a series of increasing drive strengths. The normalized
response consists of the measured voltage, $V_{a1}$, scaled by the maximum
value measured in each case.}
\label{fig:fig2}
\end{figure}

Experiments were performed on a silicon nitride beam with length $25.5$\,$\mu
$m, width $170$\,nm and thickness $170$\,nm. Metal layers consisting of
3\,nm of Ti and $40$\,nm of Au were added by thermal deposition to form a
wire on top of the beam through which a drive current could be applied. The
devices were fabricated by dry-etching in a multi-stage process\cite%
{Kunal_thesis} from wafers composed of a 190\,$\mu$m thick silicon wafer
with 170\,nm of silicon nitride on both sides; the nitride layer has a
built-in tensile stress of about $1070$\,MPa at room temperature.\cite%
{cornell}

A schematic diagram of the experimental set-up is shown in Fig.\ \ref%
{fig:fig1}a. The resonator was placed inside a $^3$He/$^4$He dilution
refrigerator and measurements were performed at $100$\,mK. The motion of the
beam was monitored using the the magnetomotive method.\cite{Cleland1999} A
magnetic field of $B=3$\,T was applied and the drive was applied by passing
an alternating current through the sample. The response to the resulting
Lorentz force was detected by measuring a voltage ($V_a$), which is related
to the input voltage at the rf pre-amplifier ($V_{0}$), as the frequency of
the drive signal was tuned through the mechanical resonance. The beam was
driven at two different frequencies by using an additional signal generator
as shown in Fig.\ \ref{fig:fig1}a. The measurements were carried out in
transmission mode, see Fig.\ \ref{fig:fig1}b, capturing a dip in the
conductance of the beam in the spectral domain.

The motion of the resonator in the magnetic field generates an emf,\cite%
{Cleland1999} which itself is related to the mechanical motion. Provided
that the resonances are well-resolved, $|\omega _{0m}-\omega _{0n}|\gg \frac{%
\omega _{0m}}{Q_{m}}+\frac{\omega _{0n}}{Q_{n}}$, for $m\neq n$, the motion
at a given frequency can be attributed to a single mode and
\begin{equation}
V_{\mathrm{emf}}=w\xi _{n}LB\frac{\partial {u}_{n}}{\partial {t}}.
\label{eq:emf}
\end{equation}%
We can relate the final measured voltage, $V_{an}$ to $V_{\mathrm{emf}}$,
and hence to the amplitude of a given mode, $r_{n}$, by applying Kirchoff's
law to the circuit (Fig.\ \ref{fig:fig1}b) and solving the corresponding
equations. Thus we find,
\begin{equation}
V_{an}=GV_{0}=\frac{Z_{0}G}{Z_{0}+Z_{e}}\left( V_{\mathrm{in}}-V_{\mathrm{emf%
}}\right) ,
\end{equation}%
where $G$ is a constant of proportionality that quantifies the overall
signal gain from the sample to the network analyzer, taking into account the
losses in the transmission cables as well as the gain provided by the
amplifiers. The gain in the transmission line will in fact vary slightly
with frequency (hence the gain will not be the same for different modes) and
local temperature variations inside the fridge. Mechanical resonance leads
to a dip in the voltage $V_{an}$, hence (assuming the circuit impedances are
real) we can characterize the mechanical response of a particular mode by,
\begin{eqnarray}
V_{Sn} &=&|\frac{Z_{0}G}{Z_{0}+Z_{e}}V_{\mathrm{in}}-V_{an}|  \notag \\
&=&\frac{Z_{0}}{Z_{0}+Z_{e}}GBLw\xi _{n}\omega _{n}r_{n},
\label{eq:vmeasured}
\end{eqnarray}%
which measures the size of the dip. Expanding Eq.\ \ref{eq:singlefreq} to lowest order in $r_n$ and using Eq.\ \ref{eq:vmeasured} leads to the explicit expression:
\begin{equation}
\omega_{n,1}=\omega_{n,0}+\alpha _{n}V_{Sn}^{2},  \label{eq:fiteq}
\end{equation}%
where $\alpha _{n}=\frac{3I_{nn}^{2}\lambda\omega_{n,0} D_{n}}{8(BL\omega _{n}\xi w)^{2}}$ and $D_{n}=\left( \frac{Z_{0}+Z_{e}}{%
Z_{0}}\frac{1}{G}\right) ^{2}$.

\section{Results}

We were able to detect the first three odd harmonics of the beam which had
frequencies 7.5, 22.85 and 39.28\,MHz respectively (the even modes do not
give rise to a signal which is directly measurable using the magnetomotive
method). Measurements on each individual mode were performed first,
capturing the response of the resonator as the driving frequency was swept
through the mechanical resonant frequency. The basic properties of the three
modes are summarized in Table I. The frequencies are those measured at the
lowest drive levels used. The Q-factors are the measured values at a field
of 3 T. Although the values are still rather high, the relatively strong
magnetic field used means that they are lower than the intrinsic Q-factors
of the modes\cite{Cleland1999} (much lower in the case of the $n=1$ mode\cite%
{Kunal_thesis} whose intrinsic Q-factor is $1.8\times10^6$ at 100\,mK).

The nonlinear parameters, $I_{nn}$, and the mode parameters, $\xi _{n}$, are
calculated numerically using a value of the intrinsic tension which we
estimate to be $T_{0}=1020$\thinspace MPa. This is slightly lower than the
room temperature value because of the different thermal contractions of the
silicon and silicon nitride layers in the wafer.\cite{Kunal_thesis} The
parameters, $I_{nn}$, and hence the strength of the nonlinearity in a given
mode, grow rapidly in size with the mode number $n$. The cross terms, $I_{nm}
$, which involve two different modes, also grow with the mode number, but
are much smaller for our device: $I_{13}=-1.9$, $I_{15}=-3.0$, and $%
I_{35}=-8.8$. In calculating the parameter $\lambda =(Ew^{2})/(2\rho L^{4})$
we use the Young's modulus of silicon nitride\cite{thesis80} $E_{SiN}=211$
GPa and neglect the contribution from the gold as it is under much less
tension and its Young's modulus, $E_{Au}=78$ GPa, is smaller. We include the
effect of the gold layer on the mass by using an appropriate average value
for the density. Using these parameters we obtain theoretical estimates of
the three mode frequencies of $7.8$, $24.2$ and $42.5$\thinspace MHz
respectively which are all close to the measured values (given in Table 1).
The discrepancies between estimated and measured frequencies arise from the
uncertainties in the device dimensions ($L\sim \pm 1\%$, $w\sim \pm 5\%$ and
$t_{Au}\sim \pm 10\%$).

\begin{figure}[htbp]
\centering{\ \includegraphics[width=6.9cm]{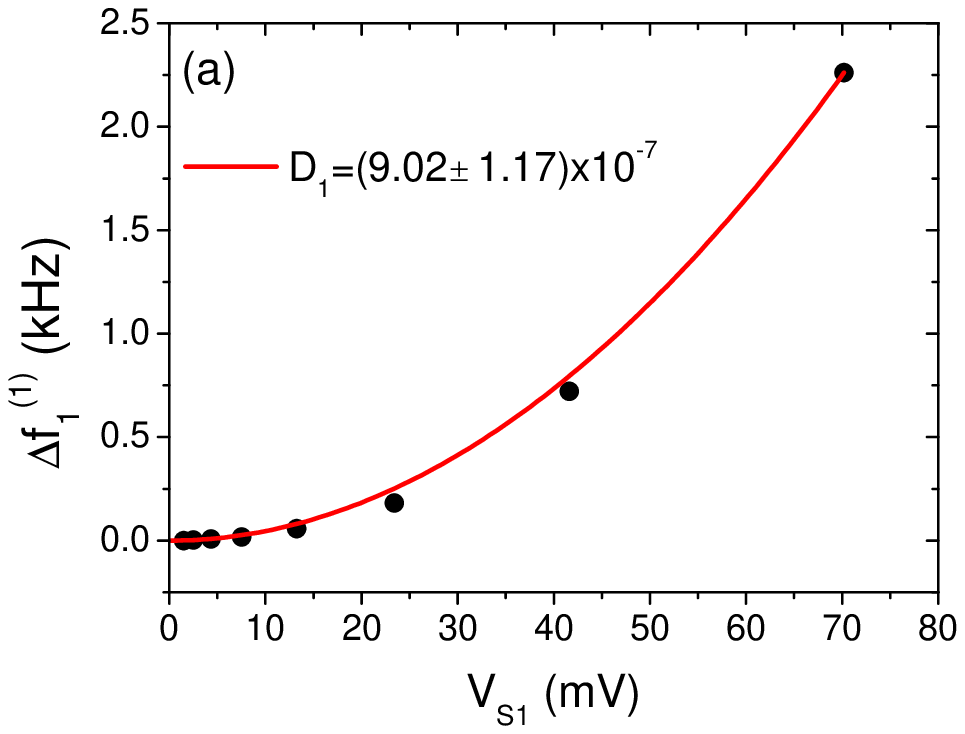} %
\includegraphics[width=6.9cm]{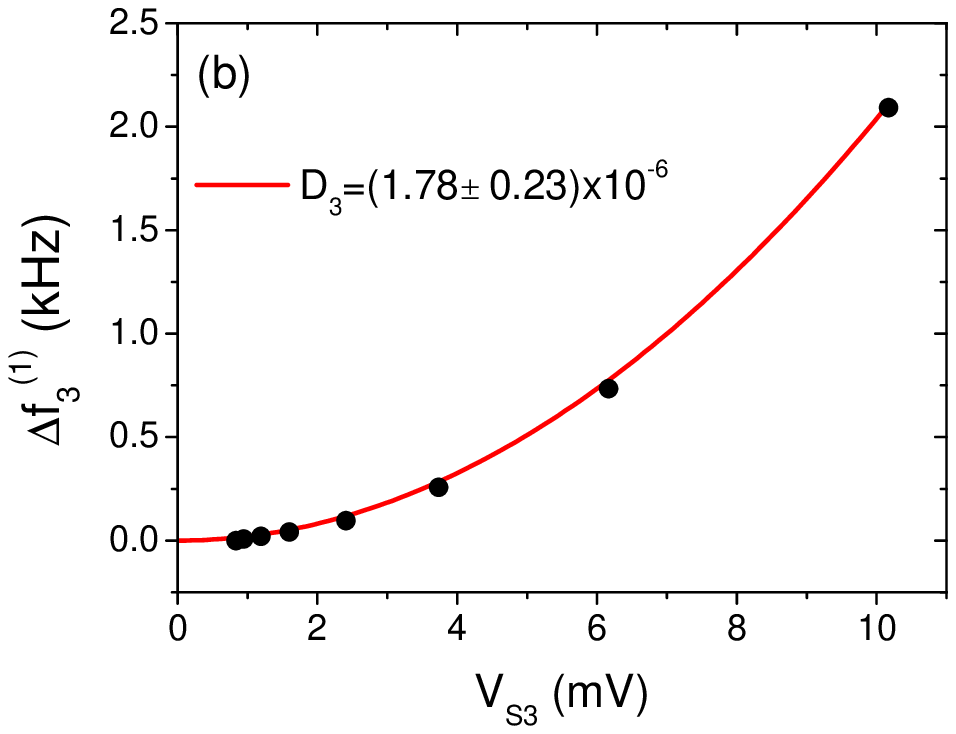} %
\includegraphics[width=6.9cm]{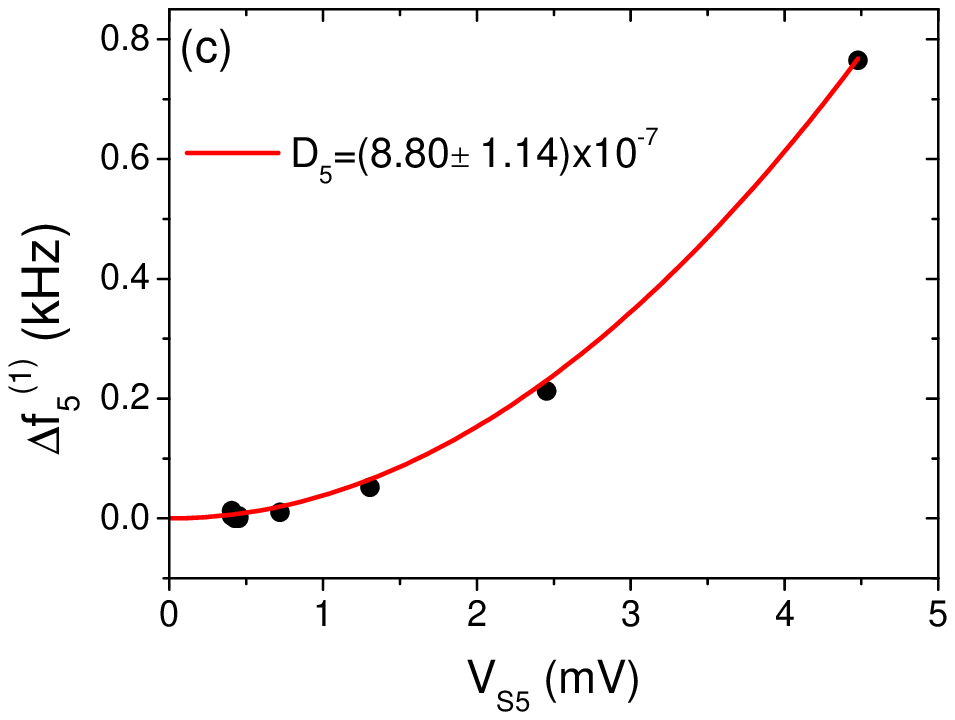}}
\caption{Shift in the frequency, $\Delta f_n^{(1)}$, as a function of the
voltage $V_{Sn}$ for (a) $n=1$, (b) $n=3$, and (c) $n=5$. The points are the
data and the lines are quadratic fits. The corresponding values of the
parameters $D_n$ obtained from the fit, together with the associated
uncertainty, are given in each case.}
\label{fig:fig3}
\end{figure}

\begin{table}[htbp]
\centering
\begin{tabular}{|c|c|c|c|c|}
\hline
\emph{n} & $\omega_{0n}/2\pi$ (MHz) & $\xi_n$ & $I_{nn}$ & $Q_n$ \\ \hline
1 & 7.50 & 0.88 & 10.4 & $9.0\times 10^3$ \\
3 & 22.85 & 0.30 & 93.3 & $1.7\times10^5$ \\
5 & 39.28 & 0.19 & 257.9 & $4.1\times10^5$ \\ \hline
\end{tabular}
\label{tab:device}
\caption{Properties of the three measured modes of the silicon nitride beam.}
\end{table}

\begin{figure}[htbp]
\centering
\includegraphics[width=8cm]{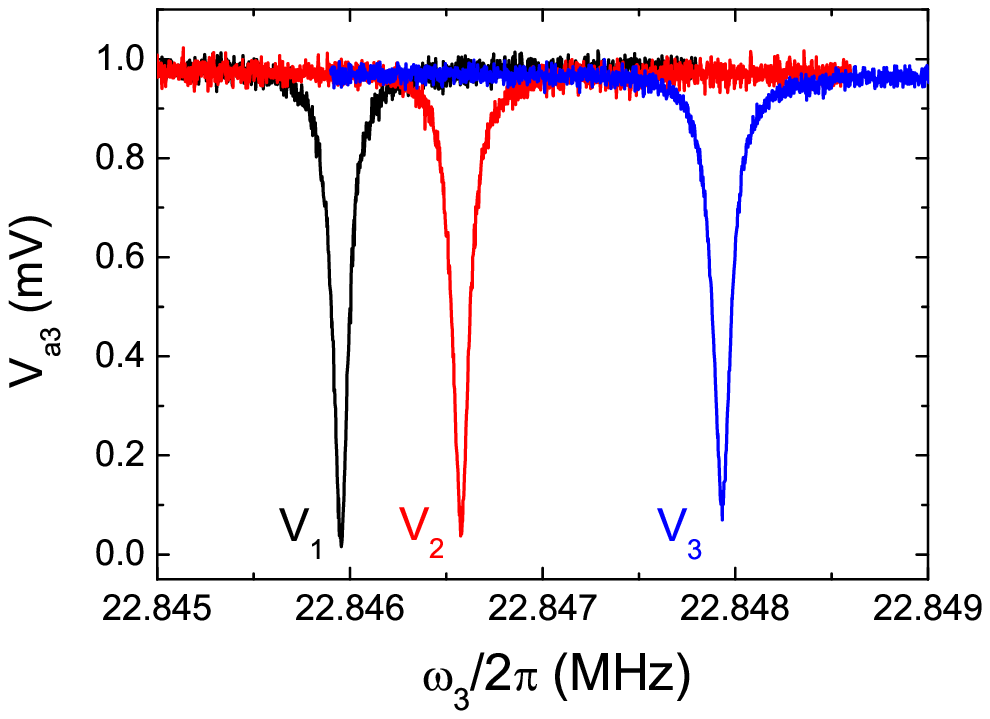}
\caption{Response curves of the third flexural mode for different measured
voltages, $V_{S1}$, at the frequency of the first mode, $V_1=15.2$ mV, $%
V_2=29.3$ mV and $V_3=48$ mV, corresponding to three successively increasing
drive amplitudes applied at a fixed frequency of 7.499 MHz.}
\label{fig:fig4}
\end{figure}

Data were collected for a range of drive amplitudes by systematically
increasing the drive signal. As an example, the spectral response of the
fundamental mode is shown in Fig.\ \ref{fig:fig2} as a function of the
corresponding drive amplitude. As expected, the curves are symmetric for the
smallest drives, but become increasingly asymmetric as the beam is driven
harder. At the largest amplitudes, the device enters a strongly nonlinear
regime marked by sharp changes seen in the measured signal on the high
frequency side of the resonance.\cite{foot1} Frequency pulling is also
clearly visible, with the peak frequency of each mode shifting upwards as
the drive is increased. Very similar behavior is seen for modes 3 and 5.

\begin{figure}[t!]
\centering
\includegraphics[width=6.9cm]{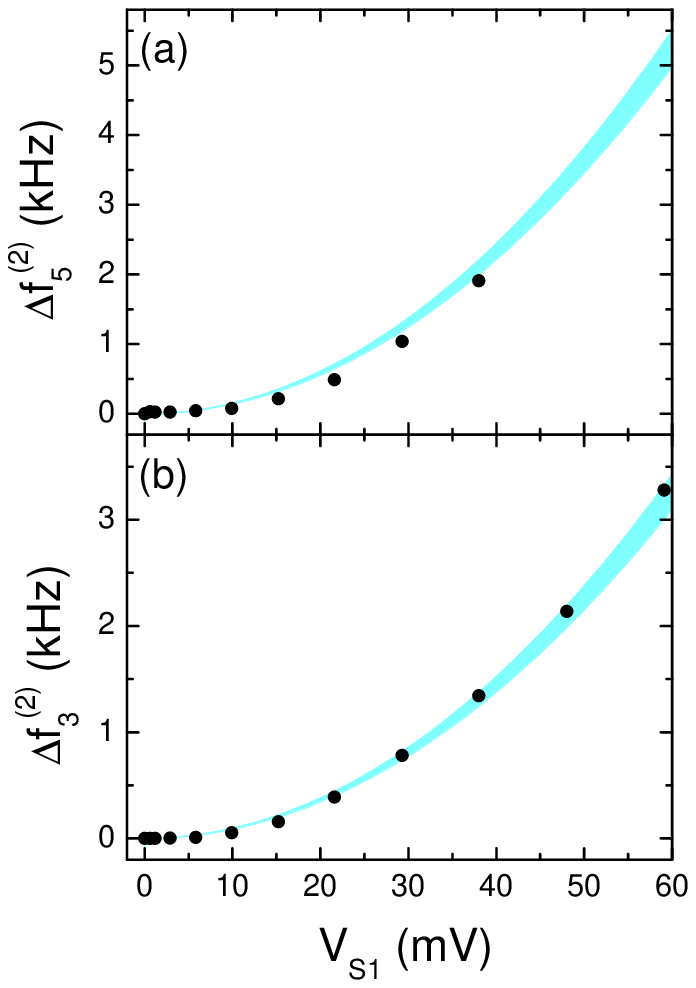}
\caption{Frequency shifts of (a) the fifth mode, $\Delta f^{(2)}_5$, and (b)
the third mode $\Delta f^{(2)}_3$, as a function of the measured response of
the first mode, $V_{S1}$, to a sequence of progressively stronger drives. In
each case the points are data, and the shaded area is the band of values consistent with theoretical predictions when uncertainties
in the parameters are accounted for.}
\label{fig:fig5}
\end{figure}

\begin{figure}[t!]
\centering
\includegraphics[width=6.9cm]{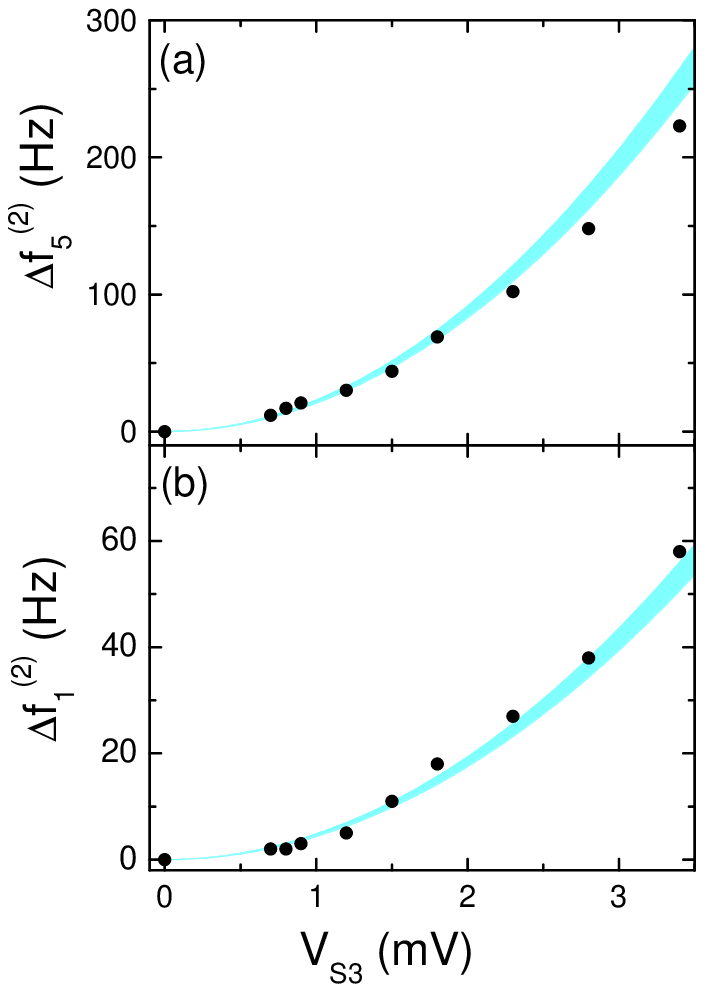}
\caption{Frequency shifts of (a) the fifth mode, $\Delta f^{(2)}_5$, and (b)
the first mode $\Delta f^{(2)}_1$, as a function of the measured response of
the third mode, $V_{S3}$, to a sequence of progressively stronger drives. In
each case the points are data, and the shaded area is the band of values consistent with theoretical predictions when uncertainties 
in the parameters are accounted for.}
\label{fig:fig6}
\end{figure}

We extracted the frequencies of peak response, $f_{n}^{(1)}=\omega
_{n,1}/2\pi $ and the corresponding voltages $V_{Sn}$ from the measured
resonance curves (Fig.\ \ref{fig:fig2}). The shift in peak frequency, $%
\Delta f_{n}^{(1)}=(\omega _{n,1}-\omega _{n,0})/2\pi $, versus peak
measured voltage $V_{Sn}$, for each mode is shown in Fig.\ \ref{fig:fig3}.
Theory predicts (see Eq. \ref{eq:fiteq}) that the frequency shift should
increase quadratically with the voltage and this is what is found. All the
parameters required for the analysis are known (with stated uncertainty)
except for the $D_{n}$. We obtain a value of the quadratic coefficient, $\alpha _{n}$, for each mode  by
fitting each set of measured data to a quadratic function with a fitting
error of less than 3\%, from which we then calculate the $D_{n}$. The fits
are shown as lines in Fig. \ref{fig:fig3}. The value of $D_{n}$ varies
between the modes by about a factor of two, due to the intrinsic frequency
dependence of the circuit impedances and amplifier gain. The error in $D_{n}$
is about 13$\%$ and follows directly from the error in sample dimensions.

Next we probed the interactions between the different modes of the beam by
exciting one mode weakly and measuring its response as successively stronger
drive amplitudes (of a fixed frequency) were applied to a second mode. The
shift in the peak response frequency of the weakly driven mode, $\Delta
f_n^{(2)}=(\omega_{n,2}-\omega_{n,0})/2\pi$, was measured together with the
voltage at the frequency of the second mode. Response curves for mode 3,
measured for three different levels of drive applied to mode 1, is shown in
Fig.\ \ref{fig:fig4}. As expected, the resonant frequency of mode 3
increases with the measured voltage from the first mode. Looking at the
depths of the curves in Fig.\ \ref{fig:fig4}, we see that increasing the
amplitude of the mode 1 drive reduces the amplitude of the response of mode
3 slightly.\cite{Westra2010} A reduction in the peak amplitude is a natural
consequence of an increase in the frequency $\omega_{n,2}$ [see Eq.\ \ref%
{eq:rn2}], but we also find a very slight degradation in the Q-factor of the
third mode accompanies the increasing drive at the fundamental frequency.

\begin{figure}[t]
\centering
\includegraphics[width=6.9cm]{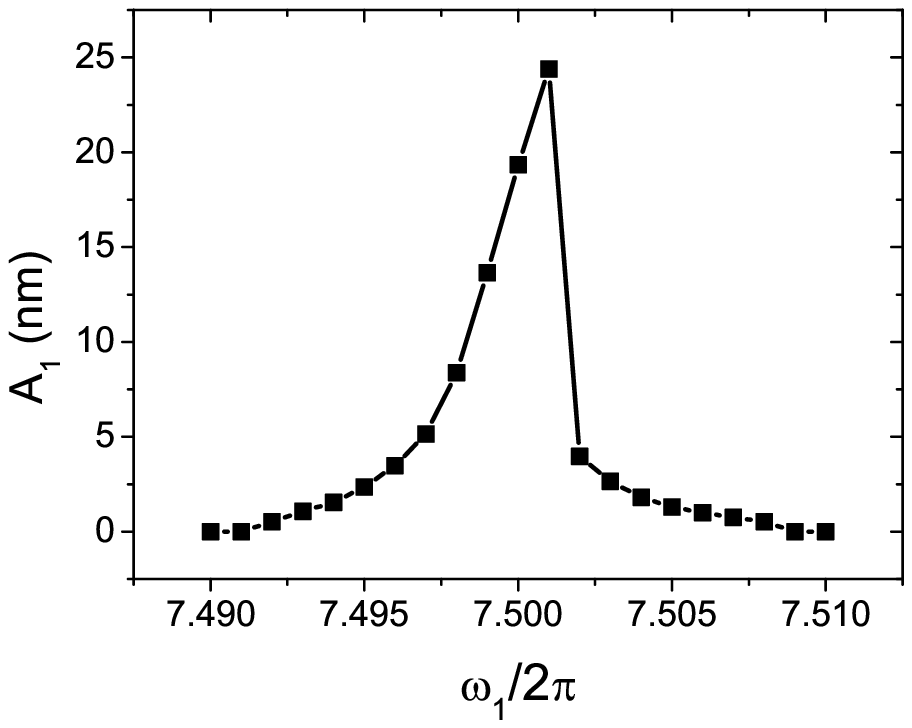}
\caption{Amplitude of the motion of mode 1 (at the antinode), $A_1$, as a
function of the drive frequency $\protect\omega_1/2\protect\pi$ obtained
using the frequency shift in mode 3. Each point is generated by measuring a
full frequency response curve for mode 3 and the obtaining the resulting $%
\Delta f^{(2)}_3 $ which is then converted into an amplitude.}
\label{fig7}
\end{figure}

The shift in the peak response frequencies of the third and fifth modes as a
function of the voltage measured at the frequency of mode 1, $V_{S1}$, are
shown in Fig.\ \ref{fig:fig5}. Similarly, Fig.\ \ref{fig:fig6} shows the
effect of varying the amplitude of the third mode on the resonant frequency
of the first and fifth modes. Again the theory predicts (through Eqs.\ \ref%
{eq:twofreq} and \ref{eq:vmeasured}) a quadratic dependence of the frequency
shifts on the measured voltages. However, because the values of the
parameters $\alpha _{n}$ have been determined from fits to the single-mode
data, the comparison with theory now involves no free parameters. We use the $\alpha_n$ values extracted from the single mode data, and the parameters $I_{nm}$, together with the associated uncertainties, to plot shaded
bands in Figs.\ \ref{fig:fig5} and \ref{fig:fig6} showing the regions which are fully consistent with the theory.  We note that the $I_{nm}$ values are quite insensitive to the changes in beam
dimensions, so the predictions of the theory are fairly precise. In each case
it is clear that the dependence of the frequency shifts on the voltages is
well described by a quadratic law, and that there is very good agreement
between the theoretical predictions and the measurements.

Having verified that the bending nonlinearity provides a quantitatively
accurate description of the modal couplings in the beam, we can now obtain
the amplitude of the motion in one mode by measuring the frequency shifts in
a second (weakly driven) mode. The theoretical expression, Eq.\ \ref%
{eq:twofreq}, and the mode function $g_n(x)$ allow us to convert from a
measured frequency shift to a physical amplitude at a given point along the
beam. An example is shown in Fig. \ref{fig7} in which the amplitude of mode
1 at the antinode $x=L/2$, obtained by measuring a frequency shift in mode
3, is shown as a function of the drive frequency, $\omega_1$. An amplitude
of 1\,nm in mode 1 translates into a frequency shift in mode 3 of $7.11$%
\,Hz, a factor of $\sim 40$ larger than that observed in Ref.\ %
\onlinecite{Westra2010}. From Fig.\ \ref{fig7} we see that an amplitude of
25\,nm ($\sim 10$\% of the width of the beam) is already well within the
non-linear regime.

\section{Conclusions}

We have measured the properties of three of the flexural modes of a highly
stressed, doubly-clamped silicon nitride nanomechanical beam. The size of
the sample, the Q-factors, the temperature and the degree of non-linearity
are all orders of magnitude different than reported by Westra et al.\,\cite{Westra2010} for an
unstressed beam at room temperature. As a first step we investigated the
frequency pulling that accompanies increases in the drive applied to each of
the modes in turn. For all three modes the frequency was found to grow
quadratically with the amplitude of the motion, consistent with the
Duffing-type behavior predicted for a nonlinearity governed by the
stretching of the beam on deflection.

We then examined the behavior of the system when one mode is driven weakly
and the amplitude of the drive applied to a second mode is increased
steadily. We again found that the frequency of the weakly driven mode
increased quadratically, this time with the amplitude of the second mode.
Using the results from the single-mode experiments to calibrate our
measurement set-up we were able to make a comparison without any free
parameters to theoretical predictions. The good agreement we find using four
different pairs of modes allows us to conclude that the nonlinear dynamics
we observe does indeed arise from the stretching of the beam on deflection.

In conclusion, our data confirm that the bending nonlinearity provides a
good quantitative description of the mode couplings in our nanomechanical
device. We find that beam starts to behave nonlinearly for amplitudes that
are a small fraction of its width, which means that the system has a high
intrinsic nonlinearity due to its size and the built-in stress in the
nitride layer. Furthermore, our measurements also allows us to calculate the
physical displacement of one mode by measuring a frequency shift in a second
mode. Studying the inter-modal couplings in nanomechanical beams is of great
importance for understanding the dynamics of such small systems, and can be
of great use to device applications or fundamental studies which require a
combination of high frequencies, large $Q$-factors and sensitivity.

\section*{Acknowledgements}

We thank E. Collin for helpful discussions and acknowledge financial support
from EPSRC (UK) under grant EP/E03442X/1.


\begin{thebibliography}{99}
\bibitem{foot1} Note that when these measurements were taken the frequency
was always swept upwards.
\end{thebibliography}
\end{document}